\documentclass[fleqn,twoside]{article}
\usepackage{espcrc2}

\usepackage{graphicx}
\usepackage[figuresright]{rotating}

\usepackage{xspace}
\usepackage{relsize}
\def\babar{\mbox{\slshape B\kern-0.1em{\smaller A}\kern-0.1em
    B\kern-0.1em{\smaller A\kern-0.2em R}}\xspace}
\def\belle{\mbox{\normalfont Belle}\xspace}
\def\cleo{\mbox{\normalfont CLEO}\xspace}
\def\cdf{\mbox{\normalfont CDF}\xspace}
\newcommand{\dzero}{D\O\xspace}
\def\atlas{\mbox{\normalfont ATLAS}\xspace}
\def\cms{\mbox{\normalfont CMS}\xspace}
\def\lhcb{\mbox{\normalfont LHCb}\xspace}

\title{Beauty 2006 -- Conference Summary and Future Prospects}
\author{Tim Gershon\address{Department of Physics, University of Warwick, Coventry, CV4 7AL, UK}}
       
\begin{document}

\begin{abstract}
The status of $B$ physics, $CP$ violation and related measurements 
at the time of the Beauty 2006 conference are summarized.  
Particular attention is given to the exciting prospects
that lie ahead, at the commencement of the LHC era, and beyond.
\vspace{1pc}
\end{abstract}

\maketitle

\section{INTRODUCTION}

Beauty 2006, 
The 11$^{th}$ International Conference on $B$ Physics at Hadron Machines,
took place at a particularly pertinent time for the subject of the conference.
While progress in $B$ physics has, in recent years, been dominated by
the spectacular successes of the $e^+e^-$ $B$ factories
(principally \babar\ and \belle, but also \cleo),
2006 can make a strong claim to be the first year for a long time 
in which the most interesting results have originated from hadron machines.
With the LHC start up becoming tantalizingly close,
does this mark the beginning of a new era for flavour physics?
Will results from \cdf\ and \dzero, and then \atlas, \cms\ and (particularly)
\lhcb\ take the headlines in the coming years,
or can the $B$ factories continue to find innovative methods to 
maximize the physics return from their unprecedented samples of data?
Looking further ahead, will an upgraded \lhcb\ be able to 
harness the enormous quantities of $b$ hadrons produced~\cite{muheim},
and fully exploit the potential for flavour physics at a hadron machine?
Will $e^+e^-$ machines reach to even higher luminosities~\cite{bevan},
with a ``Super Flavour Factory'' to complete the picture?

Such questions provided the subtext to the conference,
which was illuminated by numerous excellent presentations and 
lively discussions, conducted in a notably convivial atmosphere.
Since it appears impossible to do justice to all the content,
this summary will be selective, and focus on physics with $B$ mesons
(regrettably excluding important topics in charm~\cite{asner,other_charm} 
and charmonia~\cite{yabsley}, amongst others, that are summarized elsewhere).

\section{THE UNITARY TRIANGLE}

The agreement (or otherwise) of flavour physics results
with the Cabibbo-Kobayashi-Maskawa~\cite{ckm} mechanism for quark mixing 
is usually illustrated in terms of measurements of the properties of 
the so-called ``Unitary Triangle'' (UT).
The colourful images provided by the CKMfitter~\cite{tjampens,ckmfitter} 
and UTfit~\cite{utfit} groups clearly show the consistency (or lack thereof)
of the existing constraints with the Standard Model (SM),
to the extent that these images are nowadays almost ubiquitous
(and therefore unnecessary to reproduce here).
Below, the status of measurements of the properties of the UT 
is summarized~\cite{angles}.

\vspace{2ex}
\noindent
\underline{$\beta$} \\
\indent
The ``golden mode,'' $B^0 \to J/\psi \, K^0$, and its relatives,
provide a theoretically clean measure of $\sin(2\beta)$~\cite{lacker}.
As the main {\it raison d'\^etre} of the $B$ factories,
the measurements are updated regularly;
the most recent updates~\cite{babar:jpsik0,belle:jpsik0}
take advantage of the majority of a combined data sample that 
now exceeds $1 \ {\rm ab}^{-1}$ (about $10^9$ $B\bar{B}$ pairs).
[As Prof.~Peach imparted after the conference banquet,
the luminosity frontier is now charting the ``Attoworld'',
just as the energy frontier will soon explore the ``Terascale''.]
The world average is $\sin(2\beta) = 0.675 \pm 0.026$.

Three different approaches to resolve the ambiguity in the solutions for 
$\beta$ from the above constraint have been attempted.
They use: 
1) $B^0 \to J/\psi \, K^{*0}$~\cite{babar:jpsikstar,belle:jpsikstar};
2) $B^0 \to D^{(*)}h^0$ with $D \to K^0_S \pi^+\pi^-$~\cite{babar:dh0,belle:dh0};
and, new this year,
3) $B^0 \to D^{*+}D^{*-}K^0_S$~\cite{babar:dstardstarks}.
The results from all three prefer the SM solution 
($\beta = (21.2\pm1.0)^\circ$).
Yet while it is straightforward to draw a qualitative conclusion,
quantifying the degree to which the alternative solution is disfavoured
is extremely difficult, since each of the above methods suffers from 
highly non-Gaussian errors, either statistical in nature 
or related to hard-to-quantify hadronic parameters.
Updated measurements (particularly for $J/\psi \, K^{*}$)
and theoretical reassessments (particularly for $D^{*}D^{*}K^0_S$) will help.

\begin{figure}[htb]
  \includegraphics[width=\columnwidth]{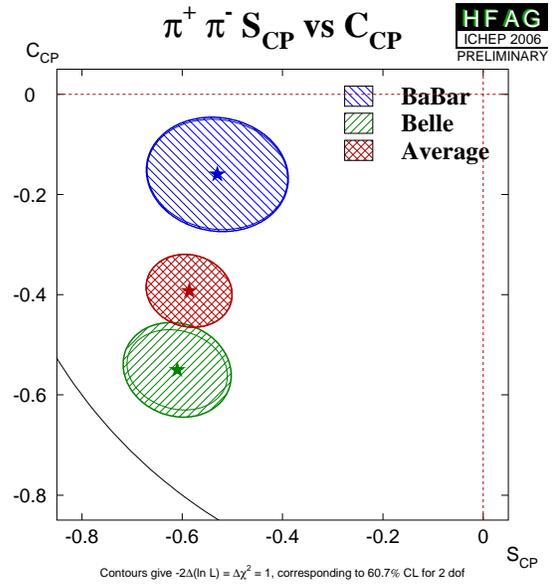}
  \vspace{-8ex}
  \caption{
    Summary of time-dependent $CP$ violation in $B^0 \to \pi^+\pi^-$.  
    $S_{CP}$ and $C_{CP}$
    are parameters of mixing-induced and direct $CP$ violation, respectively.
    The contours indicate $\Delta \chi^2 = 1$.
    For more details, see~\cite{hfag}.
  }
  \label{fig:hfag:pipi}
\end{figure}

\vspace{2ex}
\noindent
\underline{$\alpha$} \\
\indent
Measurements of $\alpha$ using mixing-induced $CP$ violation in 
$b \to u\bar{u}d$ transitions are complicated by the possible presence
of sizeable penguin contributions~\cite{zupan}.
Nonetheless, impressive progress has been made~\cite{bianchi}.
In the most recent updates in the $B \to \pi\pi$ system,
\babar~\cite{babar:pipi} have confirmed the large $CP$ violation
previously observed by \belle, who in turn have now observed~\cite{belle:pipi}
a large direct $CP$ violation effect.
The results are summarized in Fig.~\ref{fig:hfag:pipi}.
While there is still some discrepancy between the obtained values of the 
direct $CP$ violation parameter,
it now seems clear that large ($10\%$ or greater) direct $CP$ violation 
is present in $B^0 \to \pi^+\pi^-$.
Future measurements should resolve the precise value.

The extraction of $\alpha$ from these measurements is performed
using an isospin analysis, which requires measurements 
of the rates and asymmetries of the remaining $B \to \pi\pi$ decays.
Somewhat surprisingly, there are differences in the details of the outcome
between versions of the analysis performed using frequentist~\cite{ckmfitter}
or Bayesian~\cite{utfit} statistical treatments.
Nonetheless, if one bears in mind that solutions close to $\alpha = 0$
are disfavoured for various reasons, 
both analyses agree that the SM solution ($\alpha$ close to $90^\circ$)
is consistent with the data.
The same is also true for results from the $B \to \rho\pi$ system
(in which both experiments now have results from
time-dependent Dalitz plot analyses of 
$B^0 \to \pi^+\pi^-\pi^0$~\cite{babar:pipipi0,belle:pipipi0}),
and from the $B \to \rho\rho$ system~\cite{babar:rhorho,belle:rhorho},
where the first evidence for the decay $B^0 \to \rho^0\rho^0$ 
has recently been found by \babar~\cite{babar:rho0rho0}.
It will be interesting to see how well LHCb can contribute to these
challenging measurements~\cite{robbe}.

\begin{figure}[htb]
  \includegraphics[width=\columnwidth]{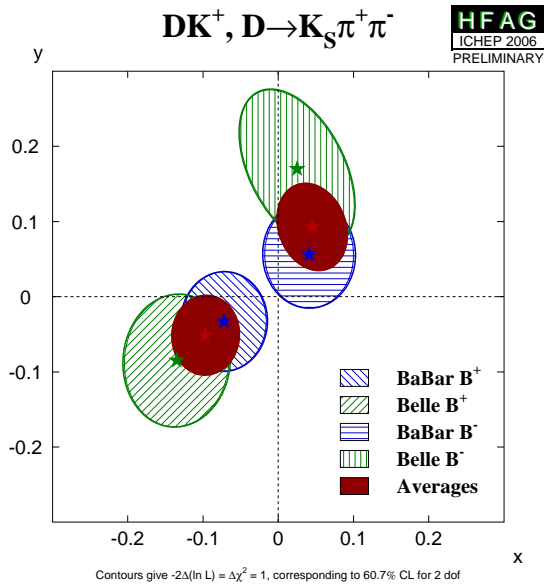}
  \vspace{-8ex}
  \caption{
    Summary of results in $B^\pm \to DK^\pm$, 
    with $D \to K^0_S\pi^+\pi^-$. 
    $CP$ violation (due to $\gamma \neq 0$) would result in 
    a difference between $B^+$ and $B^-$.
    The contours indicate $\Delta \chi^2 = 1$,
    but do not include $D$ decay model uncertainty.
    For more details, see~\cite{hfag}.    
  }
  \label{fig:hfag:dkdalitz}
\end{figure}

\vspace{2ex}
\noindent
\underline{$\gamma$} \\
\indent
The cleanest measurement of $\gamma$ can be made in the 
$B \to D^{(*)}K^{(*)}$ system~\cite{zupan},
and the most constraining results to date use 
$D \to K^0_S\pi^+\pi^-$ decays~\cite{trabelsi}.
The most recent results from \babar~\cite{babar:dkdalitz}
and \belle~\cite{belle:dkdalitz}
are summarized in Fig.~\ref{fig:hfag:dkdalitz}.
Despite impressive improvements in the analyses
(notably in the $\chi^2/{\rm ndf}$ obtained by \babar~\cite{babar:dkdalitz}
in the fit to flavour-tagged $D$ decays, used to obtain the $D$ decay model),
a large uncertainty remains due to the chosen model and hence
assumed phase variation across the Dalitz plane.
It will be very hard to significantly reduce this error without 
the concurrent analysis of large samples of $CP$-tagged $D$ mesons, 
such as those produced by CLEO-c~\cite{zupan,asner}.

To optimize the precision on $\gamma$, it is necessary to combine 
results using as many $D$ decays as possible 
(within the $B \to DK$ system)~\cite{zito}.
This year, the first analysis using $D \to K^\pm\pi^\mp\pi^0$
has been performed by \babar~\cite{babar:dkkpipip0},
which can now be added to the list of modes available to make the constraints.
Additionally, many new channels may become accessible in the coming years,
including several which are particularly interesting for analysis 
at LHCb~\cite{xie}.
Notably, none of the channels used to date have yet shown any significant
effect of the suppressed amplitude,
suggesting that the ratio of amplitudes $r_B$ 
may be smaller than na\"\i vely expected.

\vspace{2ex}
\noindent
\underline{$V_{ub}$ and $V_{cb}$} \\
\indent
The sides of the UT are obtained through measurements of rates, 
appropriately normalized.
Measurements of both $V_{ub}$ and $V_{cb}$ have reached an impressive precision,
the latter running into theoretical uncertainties 
at the $1$--$2\%$ level in inclusive channels 
(with larger theory errors for exclusive modes)~\cite{paz}.
Analysis of moments of the inclusive decay spectra 
allow continuing refinements~\cite{barberio}.
The improved understanding also aids the extraction of $V_{ub}$,
where the errors from exclusive and inclusive approaches are comparable 
(though the results are not in perfect agreement)
at about $7\%$~\cite{paz,gibbons}.
The inclusive analysis is dominated by new results from 
\babar~\cite{babar:pilnu}, \belle~\cite{belle:pilnu} 
and \cleo~\cite{cleo:pilnu} in the $B \to \pi l \nu$ channel,
complemented by improved lattice calculations~\cite{davies}.

\vspace{2ex}
\noindent
\underline{$V_{td}$ and $V_{ts}$} \\
\indent
The final side of the UT can be obtained from measurements of $\Delta m_d$,
the frequency of $B^0$--$\bar{B}^0$ oscillations,
but, to keep the theoretical uncertainty under control,
more precise constraints can be obtained if the mixing rate is 
normalized to that in the $B^0_s$ system, $\Delta m_s$.
The ratio, $\Delta m_d / \Delta m_s$ gives $|V_{td}/V_{ts}|^2$
up to SU(3) breaking terms that can be calculated in lattice QCD.

\begin{figure}[htb]
  \includegraphics[width=\columnwidth]{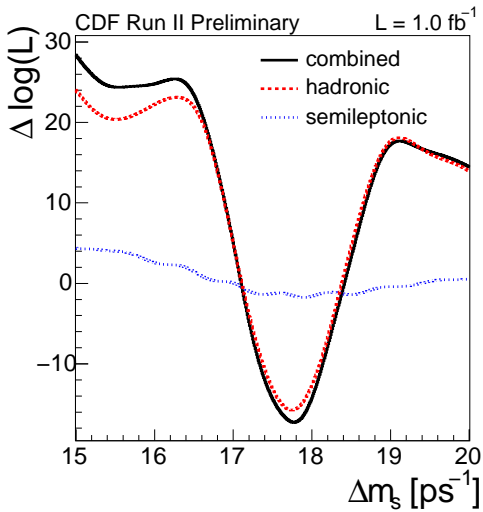}
  \vspace{-8ex}
  \caption{
    Log likelihood curves for $\Delta m_s$.
  }
  \label{fig:cdf:dms}
\end{figure}

In contrast to many measurements which are experimentally challenging 
since they rely on small effects,
$\Delta m_s$ is hard to measure since it is {\it large},
and hence resolving the $B^0_s$ oscillations is difficult.
While lower bounds on $\Delta m_s$ have existed for several years,
2006 saw the first upper bound~\cite{tamburello,d0:dms},
followed by the first measurement~\cite{cdf:dms1} of the quantity.
The sensation of Beauty 2006 was the presentation of 
the improvement of this latter analysis~\cite{belloni,cdf:dms2}, 
to include more decay modes and improved reconstruction and flavour tagging.
The significance of the oscillation signal, shown in Fig.~\ref{fig:cdf:dms},
now exceeds $5\sigma$, with the value
$\Delta m_s = 
17.77 \pm 0.10 \ ({\rm stat}) \pm 0.07 \ ({\rm syst}) \ {\rm ps}^{-1}$.
One of the many impressive features of the new analysis is the sensitivity 
(indicating the largest value of $\Delta m_s$ which could be measured) 
of $31.3 \ {\rm ps}^{-1}$ -- almost twice as large as the true value.
Another is that of the three contributions to the uncertainty on 
$|V_{td}/V_{ts}|$, that from the measurement of $\Delta m_s$ is smaller
than that from the measurement of $\Delta m_d$, 
which in turn is smaller than that from lattice calculations 
of the hadronic parameters involved~\cite{davies}.
Less than a year ago, no measurement existed;
as of Beauty 2006, $\Delta m_s$ is known to subpercent precision.
This remarkable achievement opens the door to numerous $B^0_s$ decay channels,
for which mixing-induced $CP$ violation can best be studied at hadron machines.

\section{NEW PHYSICS SEARCHES}

The above measurements, and others, combine to give constraints on 
the elements of the CKM matrix.
Since these include four free and fundamental parameters of the SM,
improving the precision of these measurements is an important goal 
in its own right.  
Yet the oft-discussed shortcomings of the SM remain,
and the paramount objective in high energy physics today is to 
search for the ``new physics'' (NP) by which at least 
some of these problems may be resolved.
$B$ physics provides a number of routes to search for NP,
not least through the consistency of UT constraints.
Indeed, there is currently ``tension'' between the 
measurements of $V_{ub}$ and $\sin(2\beta)$ discussed above,
though improvements in both experimental 
and (in the case of $V_{ub}$) theoretical uncertainties 
will be necessary to discover if this is indeed a hint of a NP signal.

\begin{figure}[htb]
  \includegraphics[width=\columnwidth]{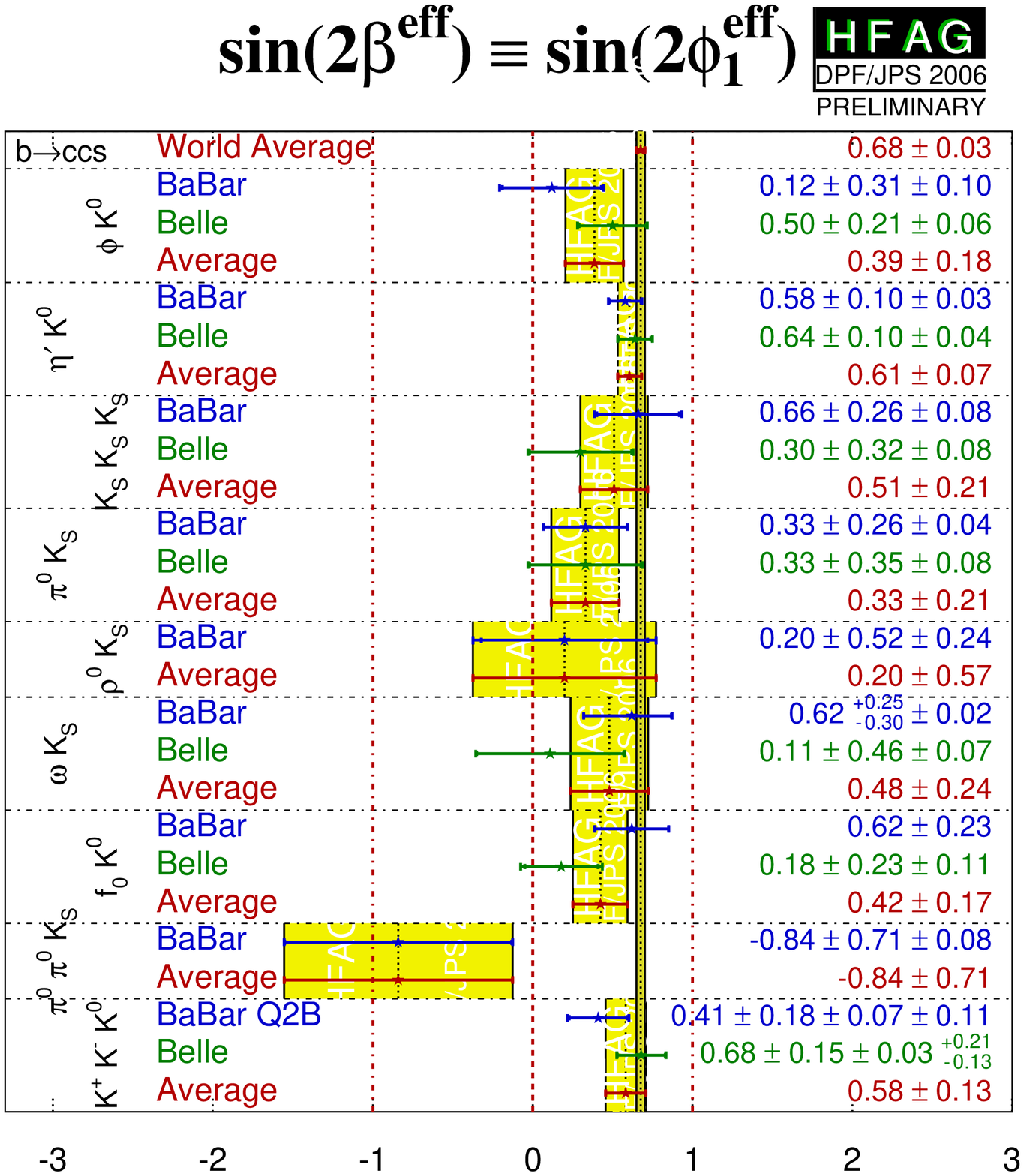}
  \vspace{-8ex}
  \caption{
    Compilation of measurements of mixing-induced $CP$ violation
    in decays dominated by the $b \to s$ penguin amplitude,
    compared to the world average from $b \to c\bar{c}s$ transitions.
    From~\cite{hfag}.
  }
  \label{fig:hfag:sPeng}
\end{figure}

\vspace{2ex}
\noindent
\underline{$\sin(2\beta^{\rm eff})$} \\
\indent
Another interesting approach to search for NP effects is to compare 
the SM reference measurement of $\sin(2\beta)$ 
with the values for the same parameter obtained in decays dominated by 
$b \to s$ penguin amplitudes~\cite{ushiroda}.
Such flavour changing neutral current transitions
are susceptible to the effects of NP particles,
which may appear as virtual particles in loops,
even if they are too massive to be observed at the energy frontier.
A compilation of relevant results is shown in Fig.~\ref{fig:hfag:sPeng}.
An important improvement was made in 2006 with the first 
time-dependent Dalitz plot analysis of $B^0 \to K^+K^-K^0$,
including intermediate states such as $\phi K^0$~\cite{babar:kkks}.
The results also include the most recent updates in the channel 
$B^0 \to \eta^\prime K^0$ from \babar~\cite{babar:etaprimeKS}
and \belle~\cite{belle:jpsik0}, wherein both
experiments have now observed (with more than $5\sigma$ significance) 
mixing-induced $CP$ violation.

It has been frequently commented upon that all the measurements 
in Fig.~\ref{fig:hfag:sPeng} take central values below the SM reference point
(meanwhile, calculations of corrections 
due to subleading SM amplitudes tend to prefer larger values).
Since each channel has different SM uncertainties,
and since there may be systematic correlations between the measurements,
taking a simple average is ill-advised (if one does so anyway, 
the significance of the effect is about $2.6\sigma$).
Moreover, there is no significant discrepancy in any particular channel.
To interpret the data therefore requires care.
Perhaps wisdom can be found in the words of Sir Francis Bacon,
whose connection with Oxford was presented by Prof.~Cashmore in the 
opening address to the conference:
\begin{quote}
  ``The root of all superstition is that men observe when a thing hits,
  but not when it misses.''
\end{quote}
The less philosophically inclined may simply find it prudent 
to wait for more data.

\begin{figure}[htb]
  \includegraphics[width=\columnwidth]{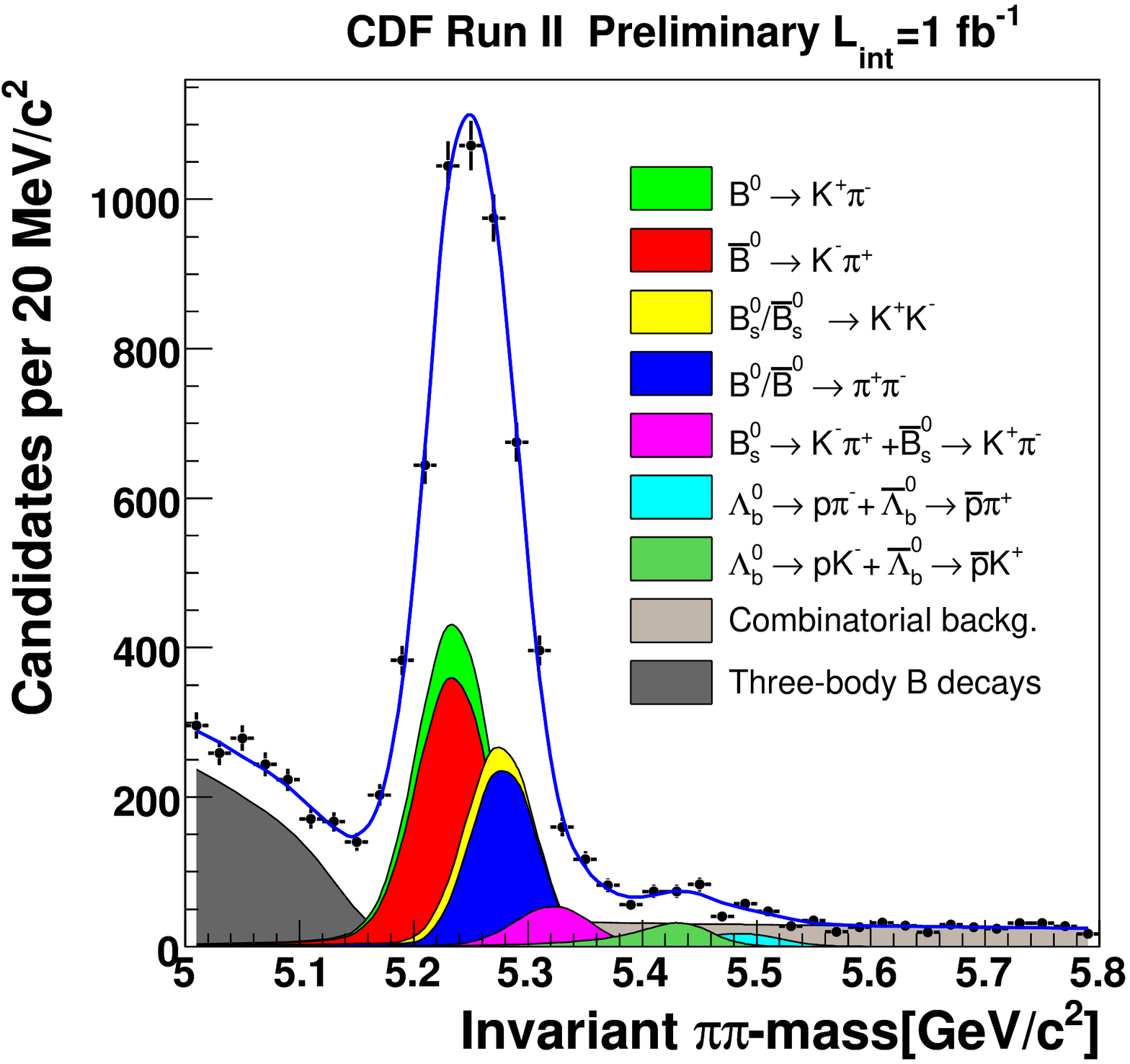}
  \vspace{-8ex}
  \caption{
    Signals for $b$ decays to $hh^\prime$ final states.
  }
  \label{fig:cdf:hh}
\end{figure}

\vspace{2ex}
\noindent
\underline{Charmless hadronic $B$ decays} \\
\indent
Decays of the type $B \to hh^\prime$ can be similarly sensitive to NP effects.
These modes have also proved fertile ground for 
testing theoretical concepts~\cite{beneke}.
For several years the data have presented a ``$K\pi$ puzzle,''
in that the pattern of rates and asymmetries in such decays was not in
good agreement with the SM prediction.
While this effect is largely reduced after the 
most recent updates~\cite{babar:hh,belle:hh},
which use improved treatments of radiative corrections,
several discrepancies remain:
the $B^0 \to \pi^0\pi^0$ branching fraction is larger than most predictions,
and most models have difficulty explaining the observed difference 
in $CP$ asymmetries for the channels $B^0 \to K^+\pi^-$ and $B^+ \to K^+\pi^0$.
Recently, results from the Tevatron have extended the experimental reach
to include $B^0_s$ and $b$ baryon decays --
first observations of three such decay channels were presented at 
Beauty 2006~\cite{morello,cdf:hh}, as shown in Fig.~\ref{fig:cdf:hh}.
These results, and their future improvements~\cite{carbone}, 
may throw new light on the ``$K\pi$ puzzle.''
However, to complete the set of measurements will require studies of decays
such as $B^0_s \to K^0_S K^0_S$ and $B^0_s \to K^0_S\pi^0$ that are not 
easily, if at all, accessible at a hadron machine, 
but may be measured at an $e^+e^-$ machine 
operating at the $\Upsilon(5{\rm S})$~\cite{blusk,y5s}.

\vspace{2ex}
\noindent
\underline{Electroweak penguin decays} \\
\indent
Since decays to hadronic final states are limited by theoretical uncertainties,
electroweak penguin decays ($b \to s\gamma$ and $b \to s ll$)
provide cleaner tests of the SM.
The measurement of the rate of the $b \to s\gamma$ decay
has provided a strict constraint that must be satisfied by 
new physics model builders;
recent improvements in the SM calculation to NNLL
allow this approach to be pushed even further~\cite{hurth}.
Meanwhile, measurements of asymmetries 
(such as $CP$, isospin and forward-backward asymmetries)
provide additional tests of the SM,
and some enticing hints for NP effects, 
that -- as usual -- require confirmation 
with larger data samples~\cite{richman,babar:kstarll,belle:kstarll}.
Since much larger statistics are necessary to probe the SM with precision,
it is gratifying that many of these channels 
(such as $B^0 \to K^{*0}\mu^+\mu^-$ and $B^0_s \to \phi \mu^+\mu^-$)
can be studied at hadron machines~\cite{lin,decapua}.

\vspace{2ex}
\noindent
\underline{Leptonic decays} \\
\indent
The archetypal channel for NP effects in flavour physics 
to be observed at a hadron collider is $B^0_s \to \mu^+\mu^-$.
This rare decay can be enhanced even in the Minimal Flavour Violation (MFV) 
scenario -- a principle that can be neatly encapsulated 
in the words of Sir Francis Bacon thus:
\begin{quote}
  ``God hangs the greatest weights upon the smallest wires.''
\end{quote}
[Rare kaon decays also play a pivotal r\^ole within MFV,
but are not within the remit of this summary.]
The latest upper limits on this channel are still 
more than an order of magnitude away from 
the SM expectation~\cite{lin,cdf:bstomumu,d0:bstomumu},
though it is expected that this mode 
can be observed by all of \atlas, \cms\ and \lhcb
within a few years of LHC data taking~\cite{smizanska}.
However, other leptonic decays, 
such as those involving $\tau$ leptons and/or neutrinos 
(notably the recently discovered $B^+ \to \tau^+\nu_\tau$~\cite{belle:taunu}),
that are also sensitive to NP,
can only be studied at a $B$ factory~\cite{ikado}.

\vspace{2ex}
\noindent
\underline{$\phi_s$ and other mixing parameters} \\
\indent
Though the value of $\Delta m_s$ (normalized to $\Delta m_d$)
has now been found to be consistent with the SM expectation,
it remains possible that there may be large NP effects 
in $B^0_s$ mixing~\cite{nierste}.
These may be uncovered through measurements of parameters such as 
$\Delta \Gamma_s$,
as well as through $CP$ violation in mixing 
({\it i.e.} the semileptonic decay asymmetry~\cite{cheu,d0:bs_asl})
and in the mixing phase $\phi_s$.
The first {\it untagged} analysis of $\phi_s$
has been carried out~\cite{chandra,d0:phis} --
note that this approach relies heavily on the size of $\Delta \Gamma_s$,
and could be pursued at the $\Upsilon(5{\rm S})$~\cite{blusk}.
However, now that $\Delta m_s$ is measured,
the complete tagged, time-dependent analysis is possible,
which will improve the sensitivity dramatically~\cite{magini}.

\section{\boldmath WHAT REMAINS FOR $B$ PHYSICS?}

The above discussion should emphasise the 
rich phenomenology for NP effects in the $B$ system.
Yet all results to date are consistent with the Standard Model,
and it is legitimate to ask if whether studies of such particles
are the best way to improve knowledge about fundamental physics.
Another way to approach this question, 
is to ask whether measurements in $B$ physics
may provide the kind of surprising result that can have 
a similar impact to the initial observation of $CP$ violation
in the decay $K^0_L \to \pi^+\pi^-$~\cite{ccft}?

Aside from the historical record of flavour physics
in uncovering new particles,
the answer is clear:
indeed such groundbreaking results are possible.
Any of the following, if observed,
would provide incontrovertible proof of physics beyond the SM:
inconsistent $CP$ violation phenomena 
(in, {\it e.g.} hadronic $b \to s$ penguin decays);
new flavour-changing neutral currents;
unpolarized photons emitted in radiative $B$ decays;
large $CP$ violation effects in $B^0_s$ mixing;
enhanced rare decays ({\it e.g.} $B^0_s \to \mu^+\mu^-$).
$CP$ violation in charm and 
lepton flavour violation in $\tau$ decays
are similarly potent observables.
Although clear signals for NP have not yet been observed,
the precision of the measurements does not exclude 
contributions of ${\cal O}(10\%)$ or, in many cases, much larger.
Some NP models allow such effects, 
though they may be unlikely in some others 
(which assume connections with other existing constraints).
Yet it must be remembered that new physics is {\it new},
and its effects are unknown.
Since searches for NP effects in flavour physics
are completely complementary to those that can be achieved
at the high energy frontier,
it is essential to continue to pursue this activity vigorously.

The preparations for the LHC are well advanced,
and entering an exhilarating stage as the first data come closer to reach.
Each of \atlas~\cite{burckhart,eerola,kirk}, 
\cms~\cite{buchmuller,schilling} and \lhcb~\cite{garrido,corti,rodrigues,ruiz}
is well positioned to exploit the $B$ physics potential of the early running.
Much has been learned from the recent operational experience 
of \cdf~\cite{annovi} and \dzero~\cite{bauer,moulik},
culminating in the results presented in this conference.
More than any other factor, this provides great hope that 
the passage from first data to published results may not be too arduous.
Notably, the \cdf\ trigger system~\cite{annovi}
has led to successful analyses of hadronic $b$ decays,
including the observation of new particles decaying into 
fully hadronic final states~\cite{pursley,states_other}.
However, any optimism should be tempered with caution,
since the clearest message of all is that a great deal of 
hard work lies in between the data and the results!

\section{SUMMARY}

It may be useful to reflect on the various observations 
of $CP$ violation to date.
Limiting the discussion to channels where effects
of more than $5\sigma$ significance have been seen,
these include $CP$ violation in
\begin{itemize}
  \vspace{-0.4\baselineskip}
  \addtolength{\itemsep}{-0.6\baselineskip}
\item $K^0-\bar{K}^0$ mixing ($\epsilon_K$);
\item interference between $s \to u\bar{u}d$ and $s \to d\bar{d}d$ amplitudes ($\epsilon^\prime$);
\item interference between $B^0-\bar{B}^0$ mixing and 
  \begin{itemize}
    \vspace{-0.4\baselineskip}
    \addtolength{\itemsep}{-0.2\baselineskip}
  \item $b \to c\bar{c}s$ amplitudes ($J/\psi \, K^0$);
  \item $b \to u\bar{u}d$ amplitudes ($\pi^+\pi^-$);
  \item $b \to s\bar{s}s$ amplitudes ($\eta^\prime K^0$);
  \end{itemize}
\item interference between $b \to u\bar{u}d$ and $b \to d\bar{d}d$ amplitudes ($\pi^+\pi^-$);
\item interference between $b \to s\bar{u}u$ and $b \to u\bar{u}s$ amplitudes ($K^+\pi^-$).
\end{itemize}

The consistency of all these observed effects with the CKM mechanism
demonstrates the tremendous success of the SM.
However, $CP$ violation has not yet been observed in the decays
of any charged meson, nor of any charmed particle,
nor of any baryon, nor of any lepton.
Clearly, a great deal remains to be explored,
and NP effects may be just around the corner.

The final summary of Beauty 2006 can be given 
in the words of Sir Francis Bacon:
\begin{quote}
  ``The best part of Beauty is that which no picture can express.''
\end{quote}

\section{ACKNOWLEDGEMENTS}

I wish to extend my warmest gratitude to the organizers
for offering me the tremendous opportunity 
to summarize such an exciting conference.
A conference in Oxford also provides the perfect opportunity
to acknowledge the enormous contributions to the field
of the late Prof.~Dalitz~\cite{dalitz_obit}.


\begin{thebibliography}{99}
\bibitem{muheim}
  F.~Muheim, these proceedings.
\bibitem{bevan}
  A.~Bevan, these proceedings.
\bibitem{asner}
  D.~Asner, these proceedings.
\bibitem{other_charm}
  See also the contributions of 
  H.~Band, these proceedings;
  R.~Muresan, {\it ibid.};
  B.~Reisert, {\it ibid.};
  P.~Zweber, {\it ibid.}
\bibitem{yabsley}
  B.~Yabsley, these proceeedings.
\bibitem{ckm}
  N.~Cabibbo, Phys. Rev. Lett. {\bf 10} (1963) 531; 
  M.~Kobayashi and T.~Maskawa, Prog. Th. Phys. {\bf 49} (1973) 652.
\bibitem{tjampens}
  S.T'Jampens, these proceeedings.
\bibitem{ckmfitter} 
  CKMfitter Group (J.~Charles {\it et al.}),
  Eur. Phys. J. {\bf C 41} (2005) 1.
  Updated results and plots available at: {\tt http://ckmfitter.in2p3.fr/}
\bibitem{utfit} 
  UTfit Collaboration (M.~Bona {\it et al.}),
  JHEP {\bf 0507} (2005) 028.
  Updated results and plots available at: {\tt http://www.utfit.org/}
\bibitem{hfag} 
  Heavy Flavour Averaging Group (E.~Barberio {\it et al.}), hep-ex/0603003.
  Updated results and plots available at: {\tt http://www.slac.stanford.edu/xorg/hfag/}
\bibitem{angles}
  Considering the location of the conference, the $(\alpha,\beta,\gamma)$ 
  notation for the angles of the Unitary Triangle is used.
  Recall that an alternative exists: 
  $(\alpha \equiv \phi_2 ,\, \beta \equiv \phi_1 ,\, \gamma \equiv \phi_3)$.
\bibitem{lacker}
  H.~Lacker, these proceeedings.
\bibitem{babar:jpsik0}
  B.~Aubert {\it et al.} (\babar\ Collaboration), 
  BABAR-CONF-06/036 (hep-ex/0607107).
\bibitem{belle:jpsik0}
  K.-F.~Chen, K.~Hara {\it et al.} (\belle\ Collaboration),
  Phys. Rev. Lett. {\bf 98} (2007) 031802.
\bibitem{babar:jpsikstar} 
  B.~Aubert {\it et al.} (\babar\ Collaboration), 
  Phys. Rev. {\bf D 71} (2005) 032005.
\bibitem{belle:jpsikstar} 
  R.~Itoh, Y.~Onuki, {\it et al.} (\belle\ Collaboration),
  Phys. Rev. Lett. {\bf 95} (2005) 091601.
\bibitem{babar:dh0} 
  B.~Aubert {\it et al.} (\babar\ Collaboration), 
  BABAR-CONF-06/017 (hep-ex/0607105).
\bibitem{belle:dh0} 
  P.~Krokovny {\it et al.} (\belle\ Collaboration),
  Phys. Rev. Lett. {\bf 97} (2006) 081801.
\bibitem{babar:dstardstarks} 
  B.~Aubert {\it et al.} (\babar\ Collaboration), 
  Phys. Rev. {\bf D 74} (2006) 091101.
\bibitem{zupan}
  J.~Zupan, these proceedings.
\bibitem{bianchi}
  F.~Bianchi, these proceedings.
\bibitem{babar:pipi} 
  B.~Aubert {\it et al.} (\babar\ Collaboration), 
  BABAR-CONF-06/039 (hep-ex/0607106).
\bibitem{belle:pipi}
  H.~Ishino, {\it et al.} (\belle\ Collaboration),
  BELLE-CONF-0649 (hep-ex/0608035).
\bibitem{babar:pipipi0}
  B.~Aubert {\it et al.} (\babar\ Collaboration), 
  BABAR-CONF-06/037 (hep-ex/0608002).
\bibitem{belle:pipipi0}
  K.~Abe {\it et al.} (\belle\ Collaboration),
  BELLE-CONF-650 (hep-ex/0609003).
  A.~Kusaka, C.C.~Wang {\it et al.} (\belle\ Collaboration),
  BELLE-PREPRINT-2007-4 (hep-ex/0701015).
\bibitem{babar:rhorho}
  B.~Aubert {\it et al.} (\babar\ Collaboration), 
  BABAR-CONF-06/016 (hep-ex/0607098).
\bibitem{belle:rhorho}
  See talk by A.~Somov at DPF/JPS 2006.
\bibitem{babar:rho0rho0}
  B.~Aubert {\it et al.} (\babar\ Collaboration), 
  BABAR-CONF-06/027 (hep-ex/0607097).
\bibitem{robbe}
  P.~Robbe, these proceedings.
\bibitem{trabelsi}
  K.~Trabelsi, these proceedings.
\bibitem{babar:dkdalitz}
  B.~Aubert {\it et al.} (\babar\ Collaboration), 
  BABAR-CONF-06/038 (hep-ex/0607104).
\bibitem{belle:dkdalitz}
  A.~Poluektov {\it et al.} (\belle\ Collaboration),
  Phys. Rev. {\bf D 73} (2006) 112009.
\bibitem{zito}
  M.~Zito, these proceedings.
\bibitem{babar:dkkpipip0}
  B.~Aubert {\it et al.} (\babar\ Collaboration), 
  BABAR-CONF-06/007 (hep-ex/0607065).
\bibitem{xie}
  Y.~Xie, these proceedings.
\bibitem{paz}
  G.~Paz, these proceedings.
\bibitem{barberio}
  E.~Barberio, these proceedings.
\bibitem{gibbons}
  L.~Gibbons, these proceedings.
\bibitem{babar:pilnu}
  B.~Aubert {\it et al.} (\babar\ Collaboration), 
  Phys. Rev. Lett. {\bf 97} (2006) 211801;
  BABAR-CONF-06-015 (hep-ex/0607060).
\bibitem{belle:pilnu} 
  T.~Hokuue {\it et al.} (\belle\ Collaboration),
  BELLE-PREPRINT-2006-10 (hep-ex/0604024).
  K.~Abe {\it et al.} (\belle\ Collaboration),
  BELLE-CONF-0666 (hep-ex/0610054).
\bibitem{cleo:pilnu}
  See talk by Y.~Gao at ICHEP 2006.
\bibitem{davies}
  C.~Davies, these proceedings.
\bibitem{tamburello}
  P.~Tamburello, these proceedings.
\bibitem{d0:dms}
  V.~Abazov {\it et al.} (\dzero\ Collaboration),
  Phys. Rev. Lett. {\bf 97} (2006) 021802.
\bibitem{cdf:dms1}
  A.~Abulencia {\it et al.} (\cdf\ Collaboration),
  Phys. Rev. Lett. {\bf 97} (2006) 062003.
\bibitem{belloni}
  A.~Belloni, these proceedings.
\bibitem{cdf:dms2}
  A.~Abulencia {\it et al.} (\cdf\ Collaboration),
  hep-ex/0609040.
\bibitem{ushiroda}
  Y.~Ushiroda, these proceedings.
\bibitem{babar:kkks}
  B.~Aubert {\it et al.} (\babar\ Collaboration), 
  BABAR-CONF-06/040 (hep-ex/0607112).
\bibitem{babar:etaprimeKS}
  B.~Aubert {\it et al.} (\babar\ Collaboration), 
  Phys. Rev. Lett. {\bf 98} (2007) 031801.
\bibitem{beneke}
  M.~Beneke, these proceedings.
\bibitem{babar:hh}
  B.~Aubert {\it et al.} (\babar\ Collaboration), 
  Phys. Rev. Lett. {\bf 97} (2006) 171805; 
  BABAR-PUB-06-047 (hep-ex/0608003);
  BABAR-CONF-06/030 (hep-ex/0607096);
  see also~\cite{babar:pipi}. 
\bibitem{belle:hh}
  S.-W.~Lin {\it et al.} (\belle\ Collaboration),
  BELLE-CONF-0633 (hep-ex/0608049);
  K.~Abe {\it et al.} (\belle\ Collaboration),
  BELLE-CONF-0632 (hep-ex/0609015);
  BELLE-CONF-0648 (hep-ex/0609006).
\bibitem{morello}
  M.~Morello, these proceedings.
\bibitem{cdf:hh}
  \cdf\ Note 8579.
  See also 
  A.~Abulencia {\it et al.} (\cdf\ Collaboration),
  Phys. Rev. Lett. {\bf 97} (2006) 211802.
\bibitem{carbone}
  A.~Carbone, these proceedings.
\bibitem{blusk}
  S.~Blusk, these proceedings.
\bibitem{y5s}
  M.~Artuso {\it et al.} (\cleo\ Collaboration), 
  Phys. Rev. Lett. {\bf 95} (2005) 261801;
  G.~Bonvicini {\it et al.} (\cleo\ Collaboration), 
  Phys. Rev. Lett. {\bf 96} (2006) 022002;
  G.S.~Huang {\it et al.} (\cleo\ Collaboration), 
  CLNS 06/1973 (hep-ex/0610035);
  A.~Drutskoy {\it et al.} (\belle\ Collaboration),
  BELLE-CONF-0614 (hep-ex/0608015);
  K.~Abe {\it et al.} (\belle\ Collaboration),
  BELLE-CONF-0615 (hep-ex/0610003).
\bibitem{hurth}
  T.~Hurth, these proceedings.
\bibitem{richman}
  J.~Richman, these proceedings.
\bibitem{babar:kstarll}
  B.~Aubert {\it et al.} (\babar\ Collaboration), 
  Phys. Rev. {\bf D 73} (2006) 092001. 
\bibitem{belle:kstarll}
  A.~Ishikawa {\it et al.} (\belle\ Collaboration),
  Phys. Rev. Lett. {\bf 96} (2006) 251801.
\bibitem{lin}
  C.J.~Lin, these proceedings.
\bibitem{cdf:bstomumu}
  \cdf\ Note 8176.
  See also 
  A.~Abulencia {\it et al.} (\cdf\ Collaboration),
  Phys. Rev. Lett. {\bf 95} (2005) 221805.
\bibitem{d0:bstomumu}
  \dzero\ Note 4733.
  See also
  V.M.~Abazov {\it et al.} (\dzero\ Collaboration),
  Phys. Rev. Lett. {\bf 94} (2005) 071802.
\bibitem{decapua}
  S.~De Capua, these proceedings.
\bibitem{smizanska}
  M.~Smizanska, these proceedings.
\bibitem{belle:taunu}
  K.~Ikado {\it et al.} (\belle\ Collaboration),
  Phys. Rev. Lett. {\bf 97} (2006) 251802.
\bibitem{ikado}
  K.~Ikado, these proceedings.
\bibitem{nierste}
  U.~Nierste, these proceedings.
\bibitem{cheu}
  E.~Cheu, these proceedings.
\bibitem{d0:bs_asl}
  V.~Abazov {\it et al.} (\dzero\ Collaboration),
  FERMILAB-PUB-07/005-E (hep-ex/0701007).
\bibitem{chandra}
  A.~Chandra, these proceedings.
\bibitem{d0:phis}  
  V.~Abazov {\it et al.} (\dzero\ Collaboration),
  FERMILAB-PUB-07/007-E (hep-ex/0701012).
\bibitem{magini}
  N.~Magini, these proceedings.
\bibitem{ccft}
  J.H.~Christenson, J.W.~Cronin, V.L.~Fitch and R.~Turlay,
  Phys. Rev. Lett. {\bf 13} (1964) 138.
\bibitem{burckhart}
  H.~Burckhart, these proceedings.
\bibitem{eerola}
  P.~Eerola, these proceedings.
\bibitem{kirk}
  J.~Kirk, these proceedings.
\bibitem{buchmuller}
  O.~Buchmuller, these proceedings.
\bibitem{schilling}
  F.-P.~Schilling, these proceedings.
\bibitem{garrido}
  L.~Garrido, these proceedings.
\bibitem{corti}
  G.~Corti, these proceedings.
\bibitem{rodrigues}
  E.~Rodrigues, these proceedings.
\bibitem{ruiz}
  H.~Ruiz, these proceedings.
\bibitem{annovi}
  A.~Annovi, these proceedings.
\bibitem{bauer}
  D.~Bauer, these proceedings.
\bibitem{moulik}
  T.~Moulik, these proceedings.
\bibitem{pursley}
  J.~Pursley, these proceedings.
\bibitem{states_other}
  R.~Mommsen, these proceedings.
  S.~Behari, {\it ibid.}
\bibitem{dalitz_obit}
  I.J.R.~Aitchison, F.E.~Close, A.~Gal and D.J.~Millener,
  Nucl. Phys. A {\bf 771} (2006) 8.

\end{thebibliography}
\end{document}